\documentclass[twocolumn,superscriptaddress]{revtex4-1}

\usepackage{amsmath}
\usepackage{amssymb}

\usepackage{float}
\usepackage{graphicx,color}
\usepackage[bookmarks=true,bookmarksnumbered=true,%
  colorlinks=true]{hyperref}

\usepackage{braket}
\newcommand{\tr}{\mathrm{tr}}

\def\({\left(}
\def\){\right)}

\newtheorem{Thr}{Theorem}
\newtheorem{Lmm}{Lemma}

\def\II{\mathbb I}

\begin{document}
\title{Secure random number generation from parity symmetric radiations}
\author{Toyohiro Tsurumaru}
\affiliation{
 Mitsubishi Electric Corporation, Information Technology R\&D Center,\\
5-1-1 Ofuna, Kamakura-shi, Kanagawa, 247-8501, Japan.}
\author{Toshihiko Sasaki}
\affiliation{
Photon Science Center, Graduate School of Engineering, The University of Tokyo,\\
2-11-16 Yayoi, Bunkyo-ku, Tokyo 113-8656, Japan.
}
\author{Izumi Tsutsui}
\affiliation{
Theory Center, Institute of Particle and Nuclear Studies,
High Energy Accelerator Research Organization (KEK),\\
1-1 Oho, Tsukuba, Ibaraki 305-0801 Japan.
}

\begin{abstract}
The random number generators (RNGs) are an indispensable tool in cryptography.
Of various types of RNG method, those using radiations from nuclear decays (radioactive RNG) has a relatively long history but their security has never been discussed rigorously in the literature.
In this paper we propose a new method of the radioactive RNG that admits a simple and rigorous proof of security.
The security proof is made possible here by exploiting the parity (space inversion) symmetry arising in the device, which has previously been unfocused but is generically available for a nuclide which decays by parity-conserving interactions. 
\end{abstract}

\maketitle

\section{Introduction}
\label{sec:introduction}

In information technology, random number generators (RNGs) refer in general to devices that output numbers distributed in a certain range uniformly.
If one wishes to use them for information security purposes in particular, their outputs must be secret \cite{Shannon1949}, in addition to being uniformly distributed.
Furthermore, in order for the RNG to be usable by anyone, these properties need to be guaranteed by some objective evidence.

Suppose, for instance, that one buys a dice from a not-necessarily-reliable vendor and throws it alone in a closed room.
For this process to generate a uniform distribution, he must be sure with an evidence that the dice is fair.
As for the secrecy, another evidence is necessary to ensure that the outputs are unpredictable and unknown to outside;
{\it e.g.}, even to the vendor or the manufacturer who had all the chances to tamper with the dice such that the outputs follow a certain pattern.
But how can one find an objective basis of secrecy that anyone can agree with?

Arguably, the most convincing basis of secrecy would be the laws of nature, that is, if nature assures the secrecy by law, then nothing can 
be utilized to predict the outputs.  This is precisely what we will adopt when we ensure the secrecy of our novel RNG proposed in this paper, under a reasonable set of assumptions which can in practice be verified without much difficulty.

In what follows, if the output of a given RNG is rigorously proven to be secret, we call it a {\it secure} RNG.
Throughout the paper we focus on secure RNGs.
The formal definition of the security, the so-called {\it universally composable security} \cite{BHLMO05}, will be given in Section \ref{sec:definition_security_LHL}; this is the most strict definition known at the present.

The secure RNG based on the laws of quantum mechanics is indeed possible \cite{Acin2016,Ma2016,Herrero-Collantes2017,Bierhorst2018,Stefanov2000,Rarity1994,Dynes2008,Ma2005a,Nie2014,Wayne2009,Wahl2011,Yan2014,Ren2011,Applegate2015,Furst2010}.
For example, RNGs using photons have been studied for a long time, and some of them have been strictly proven to be secure.
A common method of the single photon RNG is to use two complementary bases $+, \times$ of the polarization:
The legitimate user (henceforth, Alice) generates a single photon state having a polarization in one basis, say, the vertical polarization state $\ket{\updownarrow}$ belonging to basis $+$, and then measures it in the other, diagonally slanted $\times$ basis.
Alice adopts the measurement result as the random bits.

The major concern here is that the vendor of the light source may be an eavesdropper (henceforth, Eve).
In such a case, Eve could have tampered with the source to retain correlation with her own device, and may have access to the random bits as a result.

The security against such eavesdropper can still be argued as follows.
Being a pure state, the initial state $\ket{\updownarrow}$ cannot be entangled with outside, and thus has no correlation with Eve's device.
When the state is measured in the complementary basis $\times$, each measurement result, \rotatebox[origin=c]{45}{$\updownarrow$} or \rotatebox[origin=c]{-45}{$\updownarrow$}, occurs exactly with probability one half.
Thus the random bits are distributed uniformly, and they are uncorrelated with Eve.
Unfortunately, the single photon RNGs have practical disadvantages because the energy of the photon is minute and, accordingly, the detector must be highly sensitive. For this reason, the single photon RNGs suffer constraints for reduction both in their size and cost.

Besides single photon RNGs, there is another type of RNG methods which also exploit quantum phenomena, {\it i.e.}, those using radiations from nuclear decays \cite{Walker2001,Alkassar2005}.
In these radioactive RNG methods one detects radiations and adopts the timings of the detections as random numbers.
These methods were already studied half a century ago \cite{doi:10.1063/1.1658698}, and is actually older than the single photon RNG mentioned above.
The advantage of the radioactive RNG is that their device can be made smaller and simpler than that of single photon RNGs.
A sufficient sensitivity to the radiation can be achieved even with a small detector, since the energies of radiations are much larger than those of photons.
Indeed, radioactive RNG chips of a few square millimeters are already manufactured \cite{quantaglionWeb,quantaglionPatent}.

\begin{figure}[ht]
 \begin{center}
	\includegraphics[bb=0 30 310 310, width=0.3\linewidth]{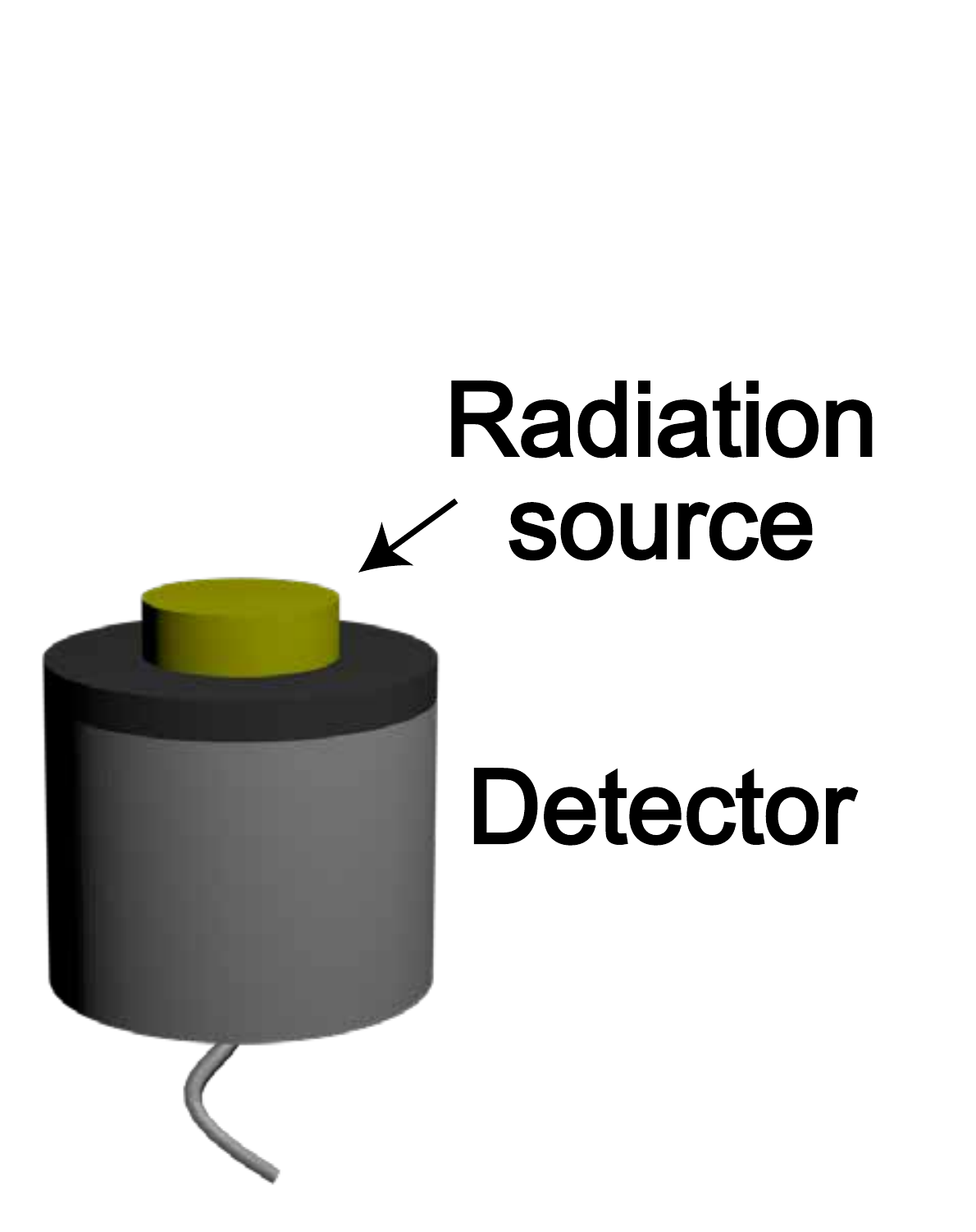}
	\caption{The device setup for the radioactive random number generator (radioactive RNG) consists of a radiation source and a detector.
We will denote the detector by $D$.
With this setup, Alice (the legitimate user) obtains raw data $\vec{i}$ by executing the procedure, steps (i), (ii) of Section \ref{sec:RNG_method}.
}
	\label{fig-actual-setup}
 \end{center}
\end{figure}

However, to the best of our knowledge, there is no security proof of the radioactive RNG, despite that 
it has been shown that they can generate a uniform distribution \cite{doi:10.1063/1.1658698}.  We find this dissatisfying, even though the concept of 
the composable security, which is essential for the proof, is relatively new \cite{BHLMO05}.

Here we present a new method of the radioactive RNG which admits a rigorous security proof.  The required security is ensured 
by the parity (space inversion) symmetry arising in the device, which is available generically for a nuclide which decays by parity-conserving interactions. 
The device structure is as simple as before, consisting only of a radiation source and a detector.
The only difference is the two conditions newly imposed on the device -- which are readily realized in practice -- which allow us to make use of the parity symmetry for ensuring security; see conditions (a) and (b) mentioned below.

The outline of our security proof is as follows.  On one hand, 
in the actual implementation, we use detection timings as the origin of randomness.
On the other hand, in the security analysis, we instead analyze the absence/presence (denoted by $z_i=0,1$) of detection in each time bin $i$, since they are merely two different formats of the same measurement results (Fig.\,\ref{fig-correspondence1}).
Then by temporarily limiting ourselves to an ideal situation (Section \ref{sec:ideal_situation} and Fig.\,\ref{fig-virtual-setup} (B)), we show that variables $z_i$ correspond to measuring the direction, up or down, of the radiation (Fig.\,\ref{fig-virtual-setup} (C)).
Hence measuring a parity symmetric radiation in this setting means measuring a parity invariant state using a pair of projectors which interchange to each other under parity operation.
Then values $z_i=0,1$ occur with an equal probability, and in addition, the resulting (sub-normalized) states on Eve's side remain fixed, irrespective of $z_i$; {\it i.e.}, Eve can gain no information of $z_i$ by any measurement.
The security in non-ideal situations can also be shown by an essentially the same argument (Section \ref{sec:realistic_situations} and Fig.\,\ref{fig-virtual-setup} (A), (D)).

\section{Main result}

\subsection{RNG method}
\label{sec:RNG_method}

\begin{figure}[t]
 \begin{center}
	\includegraphics[bb=0 0 960 500, width=\linewidth]{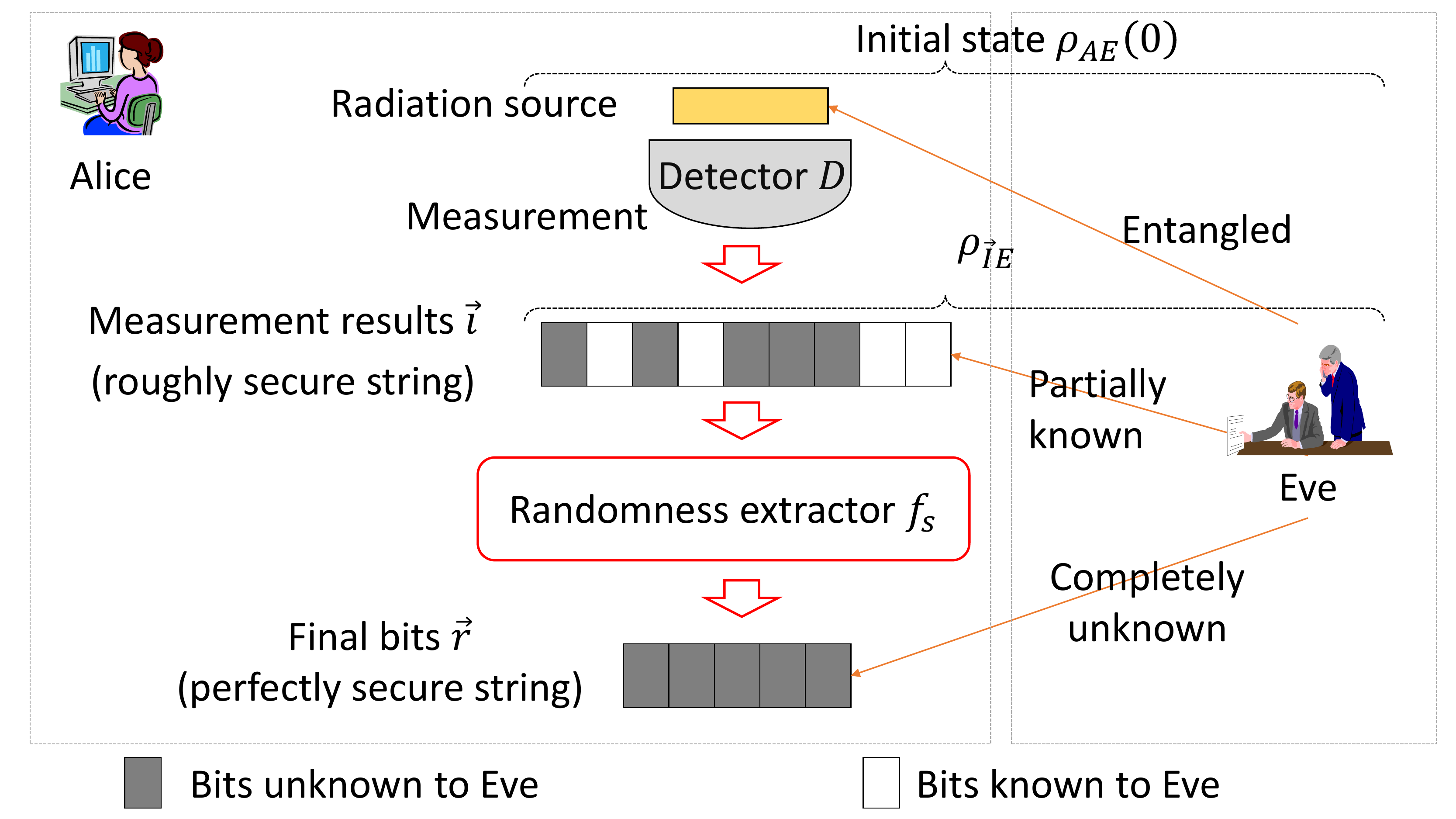}
	\caption{
The purpose of randomness extraction is to extract from a measurement result $\vec{i}$, which may be partially known to Eve, a random bits $\vec{r}$ completely unknown to Eve.
In the above picture, $\vec{i}$ being partially known to Eve is expressed by its being a mixture of black (unknown) and white (known) elements.
The number of unknown bits equals the smooth conditional min-entropy $H_{\rm min}^\delta(\vec{I}|E)$, a function of $\rho_{\vec{I}E}$.
}
	\label{fig-randomness-extraction}
 \end{center}
\end{figure}

We consider the following type of the radioactive RNG method.
By using a device consisting of a radiation source and a detector $D$ (Fig.\,\ref{fig-actual-setup}), Alice executes the following procedure (Fig.\,\ref{fig-randomness-extraction}):
Alice chooses integer parameters $N$ and $n_{\rm fin}$ such that they satisfy $0\le n_{\rm fin}\le N$.
She also selects a function $f_s$ randomly from a predetermined set of functions ${\cal F}=\{f_s\}$, each of which outputs an $n_{\rm fin}$ bit string.
Then she repeats the following steps.
\begin{description}
\item[Radioactive RNG]
\begin{itemize}
\item[(i)] 
{\bf Measurement of decay timings:}
Alice measures radiations from the source, using detector $D$, in time bins $i=1,\dots,N$.

She then records the measurement result as the list of time bins where a detection occurred;
{\it i.e.} as $\vec{i}=(i_1,\dots,i_{n_{\rm det}})$, with $n_{\rm det}$ being the number of detections, and $i_j$ being in the increasing order, $1\le i_1<i_2<\cdots<i_{n_{\rm det}}\le N$.
If there was no detection, she lets $\vec{i}=(0)$, {\it i.e.}, $n_{\rm det}=1$, $i_1=0$.
\item[(ii)]\label{step-PA}
{\bf Randomness extraction:}
Alice calculates the final bits $\vec{r}=f_s(\vec{i})$ of length $n_{\rm fin}$.
\end{itemize}
\end{description}
The purpose of each step is as follows (Fig.\,\ref{fig-randomness-extraction}).

Step (i) generates raw data $\vec{i}$ to be used as the source of the final bits $\vec{r}$.
For $\vec{r}$ to be secure, not all, but a certain fraction of $\vec{i}$ need to be unknown to Eve.
The standard theoretical results say that the size of this unknown fraction equals a quantity called the {\it smooth conditional min-entropy} $H_{\rm min}^{\delta}(\vec{I}|E)$, which is a function of the joint state $\rho_{\vec{I}E}$ of variable $\vec{i}$ and Eve (see Section \ref{sec:definition_security_LHL} and Ref. \cite{RennerPhD} for the rigorous definitions).

In step (ii) she extracts these $H^\delta_{\rm min}(\vec{I}|E)$ bits that are unknown, and generate $\vec{r}$, which is completely unknown to Eve (Section \ref{sec:definition_security_LHL} and Ref. \cite{RennerPhD}).

We denote the width of one time bin by $\Delta t$.
In order to simplify later presentations, without loss of generality, we assume that in every time bin, Alice starts her measurement at the beginning of the time bin and finishes it in a finite time $\le \Delta t$.

\subsection{Conditions on the device}
\label{sec:assumptions}

Hence the security analysis is reduced to lower bounding $H^\delta_{\rm min}(\vec{I}|E)$.
We are concerned with the possibility that the radiation source to be measured in step (i) may be entangled with Eve, and through that entanglement Eve may access $\vec{i}$;
{\it i.e.}, $H^\delta_{\rm min}(\vec{I}|E)$ may become too small to guarantee the security of $\vec{r}$ (cf. 2nd and 7th paragraphs of Section \ref{sec:introduction}).
The goal of this paper is to nullify such eavesdropping strategy by making use of the parity symmetry.

\subsubsection{Statement of conditions}
\label{sec:statement_conditions}
To this end, we assume the following three conditions on the device.
The first two of them, (a) and (b), in particular, are introduced in order to realize the parity symmetry in the device.
\begin{itemize}
\item[(a)] {\bf Radiation source:}
At the beginning of each time bin ({\it i.e.}, immediately before Alice's measurement), the state of radiations is parity invariant.
\item[(b)] {\bf Detector:} Detector $D$ is housed within one hemisphere around the source (Fig.\,\ref{fig-virtual-setup} (A)).
\item[(c)] {\bf Effect on radiations by measurements:}
Effect on radiations in the vicinity of $D$, caused by Alice's measurement of a time bin $i$, is washed away by the beginning of the next time bin $i+1$.
\end{itemize}
In addition, we introduce the following notions for later convenience.
\begin{itemize}
\item[(d)] {\bf Detections, multi-particle emissions and dark counts:}
Except with probability $\delta$, there are at least $n_{\rm thr}$ detections, at most $n_{\rm multi}$ time bins where multiple particles are emitted, and at most $n_{\rm dark}$ time bins where dark counts occur.
\end{itemize}

The statement of condition (a) requires some explanation, which we give now.
Let $\mathcal{H}_A$ be the Hilbert space describing radiated particles in the vicinity of detector $D$.
Also, let ${\cal H}_E$ be that describing all degree of freedom of Eve (cf. Fig.\,\ref{fig-randomness-extraction}).
We assume that in $\mathcal{H}_A$ the parity (space inversion) operator $P_A$ is well defined and satisfies $P_A^2=1$.
(Throughout the paper, we use the convention of omitting the identity operators included in a tensor product; hence {\it e.g.} $P_A$ is an abbreviation of $P_A\otimes 1_{E}$.)
Under this setup, we say that the joint state $\rho_{AE}(t)$ of ${\cal H}_{A}$ and ${\cal H}_{E}$ at time $t$ is parity invariant, if it satisfies
\begin{equation}
P_A\rho_{AE}(t)P_A = \rho_{AE}(t).
\label{eq:parity_invariant_state}
\end{equation}
Condition (a) says that the parity invariance (\ref{eq:parity_invariant_state}) holds at the beginning of each time bin, {\it i.e.} at $t=0,\Delta t, \dots,(N-1)\Delta t$.

\subsubsection{Feasibility of the conditions}
\label{sec:justification}
Next we discuss the feasibility of the conditions above.

First, condition (a) can basically be satisfied by choosing a nuclide which decays by parity-conserving interactions ({\it e.g.} strong and electromagnetic interactions, as in the $\alpha$- and the $\gamma$-decays), since such sources will always emit radiations with a constant parity eigenvalue.

However, as we deal here with an RNG, we must be aware of one scenario where such choice may not be sufficient for guaranteeing condition (a).
That is, the nuclide could have been tampered with by Eve, before purchased by Alice (cf. the second and seventh paragraphs of Section \ref{sec:introduction}), to the extent of destroying the parity invariance.
We point out that, even in such scenario, Alice can still verify condition (a) by performing a test on the source at hand, prior to executing the radioactive RNG.
E.g., she measures the radiation from the source and checks if the results, such as the energy spectrum and the angular distribution, are always consistent with condition (a).
If this verification succeeds she then executes the radioactive RNG; otherwise she aborts.

Second, condition (b) can always be verified visually. 

Third, condition (c) is a pure assumption.
However, this assumption is in fact implicit in most literature of quantum key distribution and physical random number generators (including the single photon RNG mentioned in Introduction).

Finally, condition (d) can be verified by statistically estimating parameters $n_{\rm thr}$, $n_{\rm multi}$ and $n_{\rm dark}$ with a significant level $\delta$, prior to executing the radioactive RNG.

\subsection{Security of measurement result $\vec{i}$}

Under these conditions, the security of measurement result $\vec{i}$ can be guaranteed as follows.
\begin{Thr}
\label{thr_min_entropy}
The smooth min-entropy $H_{\rm min}^{\delta}(\vec{I}|E)$ of $\vec{i}$, conditioned on Eve's degree of freedom $E$, is bounded as
\begin{equation}
H_{\rm min}^{\delta}(\vec{I}|E)\ge n_{\rm thr}-n_{\rm multi}-2n_{\rm dark}.
\label{eq:H_min_delta_lowerbound}
\end{equation}
\end{Thr}
This means that the final bits $\vec{r}$ are secure, if Alice chooses its length $n_{\rm fin}$ to be roughly equal to $n_{\rm thr}-n_{\rm multi}-2n_{\rm dark}$ (see Lemma \ref{lmm:rigorous_security_statement} of Section \ref{sec:definition_security_LHL} for a more rigorous interpretation of the bound (\ref{eq:H_min_delta_lowerbound})).

\section{Proof of Theorem \ref{thr_min_entropy}}
In order to simplify the analysis, we use the {\it virtual protocol} approach (also known as {\it game transform} in modern cryptography).
In this approach, instead of analyzing the actual RNG directly, one modifies it and construct a {\it virtual} RNG, as well as a quantity $H'$ arising there which lower bounds $H^\delta_{\rm min}(\vec{I}|E)$.
Then analyzing the virtual RNG, one obtains a lower bound on $H'$, which also lower bounds $H^\delta_{\rm min} (\vec{I}|E)$ by definition.
With the virtual RNG and $H'$ designed properly, this allows one to obtain a lower bound on $H^\delta_{\rm min} (\vec{I}|E)$ by a simpler analysis.

We stress that virtual RNGs will only be used for simplifying the theoretical analysis, and never need to be implemented in practice.

\begin{figure}[t]
 \includegraphics[clip, bb=0 0 900 250 clip, width=0.8\linewidth]{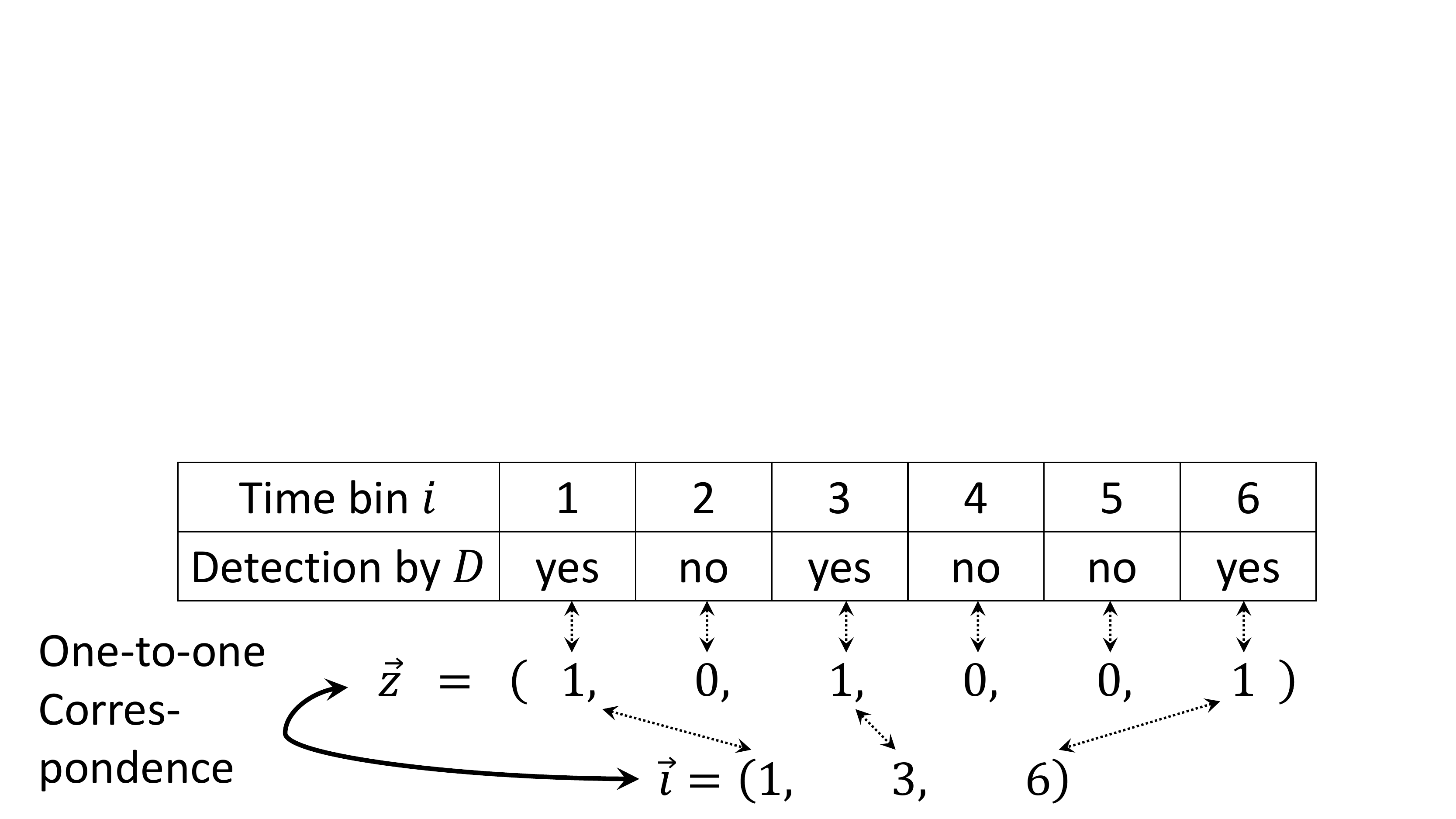}
 \caption{One-to-one correspondence between detection timings $\vec{i}=(i_1,\dots,i_{n_{\rm det}})$, and measurement results of all time bins $\vec{z}=(z_1,\dots,z_N)$.}
 \label{fig-correspondence1}
\end{figure}

As the first example of such virtual RNGs, we consider the case where Alice records the measurement result $\vec{i}$ in a different format $\vec{z}=(z_1,\dots, z_{N})$ where $z_i=0$ ($z_i=1$) indicates the absence (presence) of a detection in time bin $i$ (Fig.\,\ref{fig-correspondence1}).
In other words, Alice records measurement results $z_i$ of all time bins $i=1,\dots,N$, instead of timings $\vec{i}$ where a detection occurs.
It is straightforward to see that $\vec{i}$ and $\vec{z}$ are in a one-to-one correspondence, and are thus equally unknown to Eve,
\begin{equation}
H^\delta_{\rm min}(\vec{I}|E)=H^\delta_{\rm min}(\vec{Z}|E).
\label{eq:equality_I_Z}
\end{equation}
Thus to lower bound $H^\delta_{\rm min}(\vec{I}|E)$, it suffices to bound $H^\delta_{\rm min}(\vec{Z}|E)$; this is an example of the quantity $H'$, mentioned in the second paragraph of this subsection.

Next we will modify this virtual RNG outputting $\vec{z}$ further, such that the parity transform $P_A$, described in Section \ref{sec:assumptions}, is related to bit flips of $z_i$.
Then we will make use of this relation to lower bound $H^\delta_{\rm min}(\vec{Z}|E)$.

\subsection{Ideal situation}
\label{sec:ideal_situation}
\begin{figure}[t]
 \includegraphics[bb=0 0 960 550, width=\linewidth]{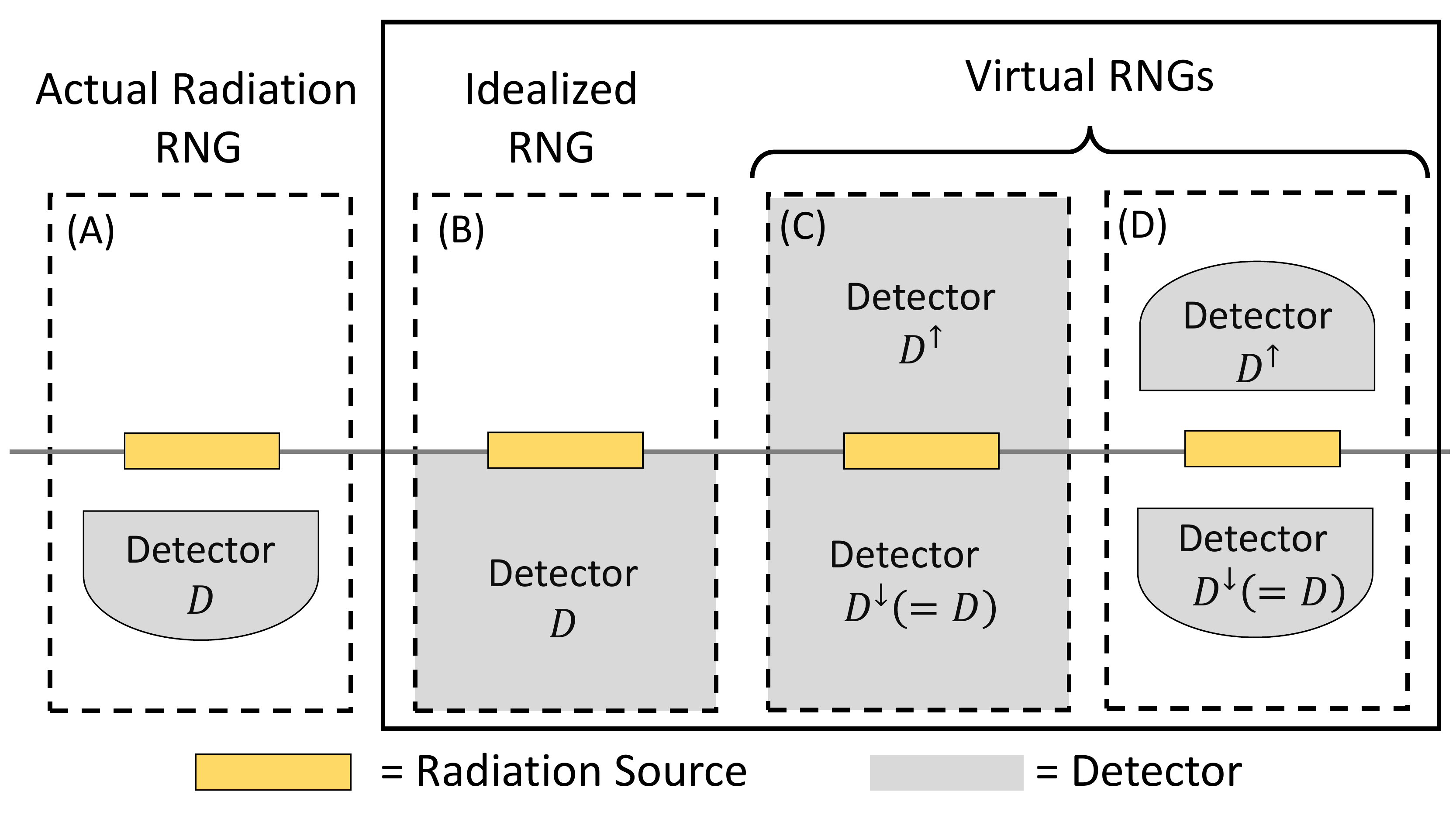}
 \caption{Item (A) is the side view of our radioactive RNG. We assume that $D$ is housed within one (the lower) hemisphere (condition (b)).
Items (B), (C) and (D) are theoretical models introduced for simplifying the description of the security proof; these three never need to be implemented in practice.
(B) is the idealized setting satisfying conditions (a') and (b'), where detector $D$ alone can determine the direction, up or down, of the particle (Section \ref{sec:ideal_situation}).
Thus (C) is equivalent to (B), the virtual RNG using two ideal detectors.
Item (D) is the the virtual RNGs corresponding to (A) (Section \ref{sec:realistic_situations}).
}
 \label{fig-virtual-setup}
\end{figure}

To elucidate this relation with a simplified situation, we temporarily idealize conditions (a) and (b) as follows.
\begin{itemize}
\item[(a')]  At the beginning of each time bin, the state of radiations is parity invariant and consists of exactly one particle.
\item[(b')]  Detector $D$ is perfect ({\it i.e.}, with a unit efficiency and no dark counts) and covers exactly the entire lower hemisphere (Fig.\,\ref{fig-virtual-setup}, (B)).
Hence $D$ goes off iff one particle or more go downward.
\end{itemize}
Then we can modify our radioactive RNG further such that bit flips of $z_i$ and $P_A$ become equivalent.

To see this, first note that detector $D$ alone can determine whether the particle went upward or downward.
Indeed, if $D$ detected the particle ($z_i=1$), it means that it went down due to (b'); and if not ($z_i=0$), two conditions together say that it went up.

These results $z_i=0,1$ can alternatively be obtained by a pair of perfect detectors, $D^\downarrow$ and $D^\uparrow$, each exactly covering the upper and the lower hemispheres (Fig.\;\ref{fig-virtual-setup}, (C)).
Thus we can define another virtual RNG satisfying (\ref{eq:equality_I_Z}).
\begin{description}
\item[Virtual RNG 1]
Using $D^\downarrow$ and $D^\uparrow$, Alice measures the source in time bins $i=1,\dots,N$, and records the result as $w_i\in\{\uparrow,\downarrow\}$.
She then lets $z_i=0,1$ if $w_i=\uparrow,\downarrow$.
\end{description}

Detectors $D^\uparrow$, $D^\downarrow$ are \lq covariant\rq\ under $P_A$;
that is, if we let $E^{\uparrow}_A$, $E^{\downarrow}_A$ be projection operators on the upper and the lower hemispheres corresponding to $D^\uparrow, D^\downarrow$, they satisfy
\begin{equation}
P_AE_A^{\uparrow}P_A=E_A^{\downarrow}.
\label{eq:covariant_E}
\end{equation}
Hence $P_A$ is equivalent to the flip of arrows $w_i=\uparrow,\downarrow$, and thus to the bit flip of $z_i$.

Next we use this parity covariance to show that $w_i$ are secure.
Recall that $\rho_{AE}$ before measurement is always parity invariant (last paragraph of Section \ref{sec:assumptions}).
Hence each $w_i$ is the result of measuring a parity invariant state $\rho_{AE}$ using parity covariant projections $E^\uparrow$, $E^\downarrow$.
Thus $w_i=\uparrow,\downarrow$ occur with an equal probability, and in addition, the resulting (sub-normalized) states on Eve's side are a fixed state, irrespective of $w_i$,
\begin{eqnarray}
\tr_{A} (E^\downarrow_A \rho_{AE})&=&\tr_{A} (P_A E^\downarrow_A P_A P_A\rho_{AE}P_A)\nonumber\\
&=&\tr_{A} (E^\uparrow_A \rho_{AE})
\label{simple-case-eq-1}
\end{eqnarray}
due to properties (\ref{eq:parity_invariant_state}) and (\ref{eq:covariant_E}).
In other words, all elements of $\vec{w}=(w_1,\dots,w_N)$ are distributed uniformly, and Eve gains no information of it by any measurement.
In terms of the min-entropy, this means
\begin{equation}
H_{\rm min}^{\delta}(\vec{Z}|E)=H_{\rm min}^{\delta}(\vec{W}|E)=N.
\end{equation}
This completes the proof of Theorem \ref{thr_min_entropy} for the ideal situation.

\subsection{General situation}
\label{sec:realistic_situations}
We proceed to the proof of the general situation. 

We again construct a virtual RNG where a correspondence between bit flips of $z_i$ and $P_A$ holds.
Alice again uses a detector pair $D^\downarrow,D^\uparrow$, with $D^\downarrow$ being the actual detector $D$  $D^\downarrow=D$) and $D^\uparrow$ being the parity transformed image of $D$ (Fig.\;\ref{fig-virtual-setup}, (D)).

As we no longer impose conditions (a') and (b'), it is possible that none or both of this detector pair, instead of one, go off in a time bin.
Hence each $w_i$ takes four values, $w_i\in\{\uparrow,\downarrow, {\rm none}, {\rm both}\}$  (Table \ref{table-correspondence2}, 1st row).

\begin{table}[t]
  \begin{tabular}{|c|c|c|c|c|} \hline
$w_i$ & $\uparrow$ & none & $\downarrow$ &both \\ \hline
$z_i=g(w_i)$ & \multicolumn{2}{|c|}{0}  & \multicolumn{2}{|c|}{1} \\ \hline
\ $\tilde{w}_i=h(w_i)$\ &\ single\ &\ none\ &\ single\ &\ both\ \\ \hline
  \end{tabular}
\caption{Relation between variables used in the proof of the general situation.
$w_i$ are outputs from detector pair $D^\downarrow, D^\uparrow$.
The output $z_i$ of the actual detector $D=(D^\downarrow)$  can be emulated from $w_i$; this corresponds to ignoring outputs of $D^\uparrow$.
$\tilde{w_i}$ denotes how many detectors went off out of $D^\downarrow$ and $D^\uparrow$.
}
\label{table-correspondence2}
\end{table}
In this case, the output $z_i$ of $D(=D^\downarrow)$ alone can be emulated from $w_i$, by ignoring outputs of $D^\uparrow$ (Table \ref{table-correspondence2}, second row).
Thus we can define a virtual RNG as,
\begin{description}
\item[Virtual RNG 2]
Using $D^\downarrow$ and $D^\uparrow$, Alice measures the source in time bins $i=1,\dots,N$, and records the result as $w_i\in\{\uparrow,\downarrow, {\rm none}, {\rm both}\}$.
She then lets $z_i=g(w_i)$, using function $f$ specified in the second row of Table \ref{table-correspondence2}.
\end{description}
whose output $g(w_i)$ satisfies
\begin{equation}
H_{\rm min}(\vec{Z}|E)=H_{\rm min}(g(\vec{W})|E).
\label{eq:H_min_ZE_eq_H_min_WE}
\end{equation}

As in the previous subsection, we can bound the right hand side of (\ref{eq:H_min_ZE_eq_H_min_WE}) by exploiting the relation between measurement results and the parity transform $P_A$.
However, the argument needs to be modified, as the relation is not the same as in the ideal situation.

That is, unlike in the ideal situation, the bit flip of $z_i$ and $P_A$ may not be equivalent in general.
This is because  $z_i=0,1$ may come from measurement results $w_i=\text{`none'}$ or $\text{`both'}$, whose quantum measurements are not in general covariant under $P_A$.
On the other hand, measurements of $w_i=\uparrow$ and $\downarrow$ are still covariant under $P_A$, by definition of $D^\downarrow, D^\uparrow$.

Hence if we evaluate the min-entropy of $w_i$ in single detection events ({\it i.e.}, time bins $i$ where $w_i=\uparrow$ or $\downarrow$; see Table \ref{table-correspondence2}, 3rd row), we have the ideal situation again, and the security can be shown by the same reasoning as in the previous subsection.
The min-entropy thus obtained lower bounds $H_{\rm min}(g(\vec{W})|E)$ on the right hand side of (\ref{eq:H_min_ZE_eq_H_min_WE}), since in general, the entropy of a part is not greater than that of the total.
As a result, $H_{\rm min}(g(\vec{W})|E)$ is lower bounded by the number of single detection events.
(For the rigorous proof of statements made in this paragraph, see Section \ref{sec:revealing_gz}.)

We can bound the number of single detection events as follows.
The number $D$ of the detection events is no larger than the sum of the number of the single detection events and the $\text{`both'}$ events.
The $\text{`both'}$ events can occur if the multiparticle emission or the dark count occurs in either detector.
Then due to condition (d), the number of single detection events can be further lower bounded by $n_{\rm thr}-n_{\rm multi}-2n_{\rm dark}$, except for probability $\delta$, and we obtain Theorem \ref{thr_min_entropy}.

\section{Methods}

\subsection{Definition of security and the leftover hashing}
\label{sec:definition_security_LHL}

We review definition of the security of RNG, as well as techniques to for guaranteeing it.

In Introduction, we said that the final bits $\vec{r}$ is secure when it is distributed uniformly and unknown to Eve.
This can be formalized as follows.
Given an actual state $\rho_{\vec{R}E}$, we define the corresponding ideal state to be $\rho_{\vec{R}E}^{\rm ideal}=2^{-n_{\rm fin}}\II_{\vec{R}}\otimes\rho_E$, $\rho_E={\rm tr}_A(\rho_{AE})$,
where $\vec{r}$ is distributed uniformly and is completely unknown to Eve.
${\cal H}_{\vec{R}}$ is the Hilbert space of the memory storing $\vec{r}$. 
However, as it is practically difficult to always guarantee this ideal situation, it is customary to relax this notion and say that $\vec{r}$ is $\varepsilon$-secure if
\begin{eqnarray}
\frac12\left\|\rho_{\vec{R}E}-\rho_{\vec{R}E}^{\rm ideal}\right\|_1\le \varepsilon,
\label{eq:average_trace_dist_defined}
\end{eqnarray}
where $\|A\|_1={\rm tr}\left(\sqrt{AA^\dagger}\right)$ denotes the $L_1$-norm of an operator $A$.
Intuitively, this says that the actual state cannot be discriminated from the ideal state except with probability $\varepsilon$.
This notion of security using parameter $\varepsilon$ is often called the universally composable security \cite{BHLMO05}.

In Section \ref{sec:RNG_method}, we stated that for the final bits $\vec{r}$ to be secure, it suffices that the smooth conditional min-entropy $H^\delta_{\rm min}(\vec{I}|E)$ of measurement results $\vec{i}$ is lower bounded.
The rigorous results corresponding to this statement are as follows.

The conditional min-entropy $H_{\min}(\vec{I}|E)_{\rho_{\vec{I}E}}$ of a sub-normalized state $\rho_{\vec{I}E}$ is defined to be the maximum real number $\lambda$, satisfying
$2^{-\lambda} \II_{\vec{I}}\otimes \sigma_E \ge \rho_{\vec{I}E}$ for a normalized state $\sigma_E$ \cite{RennerPhD,TomamichelPhD}.
We abbreviate $H_{\min}(\vec{I}|E)_{\rho_{\vec{I}E}}$ as $H_{\min}(\vec{I}|E)$, whenever the subscript $\rho_{\vec{I}E}$ is obvious from the context.
The {\it smooth} conditional min-entropy $H_{\min}^{\delta}(\vec{I}|E)_{\rho_{\vec{I}E}}$ is the maximum value of $H_{\min}(\bar{\rho}_{AE}|E)_{\bar{\rho}_{\vec{I}E}}$ of sub-normalized states $\bar{\rho}_{\vec{I}E}$ that are $\delta$-close to $\rho_{\vec{I}E}$ in terms of the purified distance
\cite{TomamichelPhD}.

If Alice performs randomness extraction (step (ii) of Section \ref{sec:RNG_method}) using a universal$_2$ function family \cite{CARTER1979143}, ${\cal F}$, the security of its output $\vec{r}$ satisfies the following.
\begin{Lmm}[Leftover hashing lemma (LHL, \cite{RennerPhD})]
\label{lmm:leftover_hashing_lemma}
If function set ${\cal F}$ is universal$_2$, and function $f_s\in {\cal F}$ is chosen with probability $p(s)$,
\begin{equation}
\sum_{s}p(s)\left\|\rho_{\vec{R}E}-\rho_{\vec{R}E}^{\rm ideal}\right\|_1
\le 2\delta+2^{\frac12[n_{\rm fin}-H^{\delta}_{\rm min}(\vec{I}|E)]}.
\label{eq:original_smoothed_leftover_hashing_lemma}
\end{equation}
\end{Lmm}

By combining this lemma and Theorem \ref{thr_min_entropy}, we can guarantee the security of $\vec{r}$ as follows.
\begin{Lmm}
\label{lmm:rigorous_security_statement}
For a given security parameter $\varepsilon>0$, the final bits $\vec{r}$ is $\varepsilon+\delta$-secure, if Alice uses a universal$_2$ hash function for randomness extraction, and if its output length $n_{\rm fin}$ satisfies
\begin{equation}
n_{\rm fin}\le n_{\rm thr}-n_{\rm multi}- 2n_{\rm dark}-2\log_2\frac1{\varepsilon}+2.
\label{eq:bound_on_n}
\end{equation}
\end{Lmm}
Recall that $n_{\rm multi}$ and $n_{\rm dark}$ depend on $\delta$ through condition (d).
Hence the right hand side of (\ref{eq:bound_on_n}) depends on both $\varepsilon$ and $\delta$.

\subsection{Detailed descriptions of Radioactive RNG and Virtual RNG 2}

We here give a detailed mathematical description of Radioactive RNG and Virtual RNG 2.
We will describe Virtual RNG 2 only, but the same description applies also to Radioactive RNG if one neglects output of virtual detector $D^\uparrow$ (cf. Table \ref{table-correspondence2}, 1st and 2nd rows).

\subsubsection{Description of the procedures of Virtual RNG 2}
\label{sec:description_Virtual_RNG2}
We will denote by $\bar{D}$ the measurements setup consisting of detector pair $D^{\uparrow}, D^{\downarrow}$.
We denote four output patterns of from $\bar{D}$ in one time bin by $w\in{\cal W}$, where ${\cal W}:=\{\uparrow,\downarrow, {\rm none}, {\rm both}\}$ (Table \ref{table-correspondence2}, 1st row).
For the convenience of the security proof, we classify $w$ by how many of the detector pair $D^{\uparrow}, D^{\downarrow}$ go off in the time bin, using symbols $\tilde{\cal W}:=\{\text{none},{\rm single}, {\rm both}\}$, where `single' event means $w=\uparrow$ or $\downarrow$.
A function $h$ can be defined corresponding to this classification (Table \ref{table-correspondence2}, third row).

We continue to describe radiated particles by the Hilbert space $\mathcal{H}_A$.
In addition, we introduce $\mathcal{H}_B$ to describe the radiation source.

We describe the quantum process (measurement and time evolution) occurring inside the RNG device, during the beginnings of adjacent time bins, by a completely positive map $M_{AB}^{w}:{\cal H}_A\otimes {\cal H}_B\to {\cal H}_A\otimes {\cal H}_B$.
That is, if Alice measures the state $\sigma_{ABE}(j\Delta t)$ at the beginning of time bin $j+1$ and obtains output $w$, the state at the beginning of next time bin is $\sigma_{ABE}^w((j+1)\Delta t)=M_{AB}^w(\sigma_{ABE}(j\Delta t))$.

(We here extend the convention for operators, introduced above eq. (\ref{eq:parity_invariant_state}), to maps of states, and omit the identity operation included in a tensor product; hence {\it e.g.} $M_{AB}^w=M_{AB}^w\otimes {\rm id}_{E}$ with ${\rm id}_{E}$ being the identity operation in ${\cal H}_{E}$.)

Hence if Alice started Virtual RNG 2 with the state $\rho_{ABE}(0)$, and measured $w_1,\dots,w_j$ in time bins $1,\dots,j$, the (sub-normalized) state at the beginning of time bin $j+1$ takes the form
\begin{equation}
\rho^{(w_1,\dots,w_j)}_{ABE}(j\Delta t):=M^{w_j}_{AB}\circ \cdots \circ M^{w_1}_{AB}(\rho_{ABE}(0)).
\label{eq:rho_z_E_multibit_defined}
\end{equation}

When Virtual RNG 2 is finished, the joint state of the memory that stores the entire measurement result $\vec{w}=(w_1,\dots,w_N)$ and of Eve takes the form
\begin{eqnarray}
\rho_{\vec{W}E}&=&\sum_{\vec{w}\in{\cal W}^N}\ket{\vec{w}}\bra{\vec{w}}_{\vec{W}}\otimes\rho^{\vec{w}}_E,\\
\rho^{\vec{w}}_E&=&\rho^{(w_1,\dots,w_N)}_E={\rm tr}_{AB}\left(\rho^{(w_1,\dots,w_N)}_{ABE}(N\Delta t)\right)
\label{eq:final_state_wE}
\end{eqnarray}

\subsubsection{Parity invariance of the measurement result $w_i$}
\label{eq:parity_invariance_w}

In this setting, we can argue that $\rho^{\vec{w}}_E$ are invariant under flips of arrows $\uparrow$ and $\downarrow$ included in $w_i$, by essentially the same argument as in Eq.\;(\ref{simple-case-eq-1}).

To see this, first note that condition (a) asserts that
\begin{equation}
 \tilde{P}_{A} (\rho^{(w_1,\dots,w_j)}_{ABE}(j\Delta t))  = \rho^{(w_1,\dots,w_j)}_{ABE}(j\Delta t).
	\label{eq:parity_invariance_rho_w}
\end{equation}

Also note that the following relation holds for maps $M_{AB}^{\uparrow}$ and $M_{AB}^{\downarrow}$,
\begin{equation}
M_{AB}^{\uparrow}\circ\tilde{P}_{A} = M_{AB}^{\downarrow},
\label{eq:parity_covariance_L_A}
\end{equation}
where $\tilde{P}_A(\rho_A):= P_A \rho_{ABE} P_A$.
Eq. (\ref{eq:parity_covariance_L_A}) holds for the following two reasons: i) Due to the construction of $\bar{D}$, obtaining the measurement result $\downarrow$ is equivalent to first applying the parity transform and then obtaining $\uparrow$. ii) Due to condition (c), the effect caused on radiations by the measurement of a time bin $i$ (which may depend on results $w_i=\downarrow, \uparrow$) is washed away before the measurement of the next time bin $i+1$ starts.

From relations (\ref{eq:parity_invariance_rho_w}), (\ref{eq:parity_covariance_L_A}), we see that the (sub-normalized) state at the beginning of time bin $j+1$ satisfies
\begin{equation}
\begin{split}
&  \rho^{(w_1,\dots,w_{j-1},\downarrow)}_{ABE}(j\Delta t)\\
=& M^{\downarrow}_{AB}(\rho^{(w_1,\dots,w_{j-1})}_{ABE}((j-1)\Delta t))\\
=& M^{\uparrow}_{AB}\circ P_{A}(\rho^{(w_1,\dots,w_{j-1})}_{ABE}((j-1)\Delta t))\\
=& M^{\uparrow}_{AB}(\rho^{(w_1,\dots,w_{j-1})}_{ABE}((j-1)\Delta t))\\
=& \rho^{(w_1,\dots,w_{j-1},\uparrow)}_{ABE}(j\Delta t).
\end{split}
\end{equation}
Further, combining this with eq. (\ref{eq:rho_z_E_multibit_defined}), we see that $\rho^{\vec{w}}_E$ are invariant under flips of arrows $\uparrow$ and $\downarrow$ included in $w_i$.
Or in terms of classification $\tilde{\cal W}=\{\text{none},{\rm single}, {\rm both}\}$
\begin{equation}
\rho^{\vec{w}}_E=\rho^{{\vec{w}}'}_E\quad {\rm if}\quad h(\vec{w})=h(\vec{w}'),
\label{eq:rho_z_E_parity_invariance}
\end{equation}
where $h(\vec{w}):=(h(w_1),\dots,h(w_N))$.
That is, $\rho^{\vec{w}}_E$, $\rho^{{\vec{w}}'}_E$ are equal, if it holds for all time bin $i$ that the number of detectors that went off in time bin $i$ is equal, $h(w_i)=h(w_i')\in\tilde{\cal W}$.

\subsection{Supplement to the proof of Theorem 1}
\label{sec:supplement_proof}
\label{sec:revealing_gz}

In the second paragraph from the last of Section \ref{sec:realistic_situations}, we argued that the right hand side of (\ref{eq:H_min_ZE_eq_H_min_WE}) is lower bounded by the number of single detection events.
The argument made there was in fact rather intuitive and not sufficiently rigorous.
Below we give a rigorous proof.

Under these settings, we consider the following virtual RNG.
This corresponds to the situation where Alice intentionally reveals $h(\vec{w})$ to Eve.
\begin{description}
\item[Virtual RNG 3] After executing Virtual RNG 2, Alice tells Eve $h(\vec{w})$.
\end{description}
The min-entropy corresponding to this case lower bounds the right hand side of (\ref{eq:H_min_ZE_eq_H_min_WE}), since Eve's ambiguity never increases on receiving an extra information $h(\vec{w})$.
\begin{equation}
H_{\rm min}(g(\vec{W})|E)\ge H_{\rm min}(g(\vec{W})|h(\vec{W}),E).
\label{eq:H_min_fZ_Hmin_fZT}
\end{equation}
After Virtual RNG 3, Alice and Eve both know the classical random variable $\vec{\tilde{w}}=h(\vec{w})$, so the overall state becomes a classical ensemble of those labeled by $\vec{\tilde{w}}$.
Thus it suffices to analyze each $\vec{\tilde{w}}$ separately.
To rephrase this rigorously, due to Lemma 3.1.8 of Ref. \cite{RennerPhD},
\begin{equation}
H_{\rm min}(g(\vec{W})|h(\vec{W}),E)
\ge \min_{\vec{\tilde{w}}}H_{\rm min}(g(\vec{W})|h(\vec{W})=\vec{\tilde{w}},E),
\label{eq:max_tH_min}
\end{equation}
where the minimum is evaluated for all values of $\vec{\tilde{w}}$ possible, {\it i.e.}, all $\vec{\tilde{w}}\in \tilde{\cal W}^N$ satisfying $\Pr(h(\vec{w})=\vec{\tilde{w}}\ |\ \rho_{\vec{W}E})>0$.

$H_{\rm min}(g(\vec{W})|h(\vec{W})=\vec{\tilde{w}},E)$ on the right hand side of (\ref{eq:max_tH_min}) measures the fraction of $g(\vec{w})$ unknown to Eve, under the restriction that $\vec{w}$ takes values satisfying $h(\vec{w})=\vec{\tilde{w}}$.
As can easily be seen by definition of functions $g$ and $h$ in Table \ref{table-correspondence2}, under this restriction, function $g$ becomes one-to-one, and thus the minimum entropies of $g(\vec{w})$ and $\vec{w}$ are equal,
\begin{equation}
H_{\rm min}(g(\vec{W})|h(\vec{W})=\vec{\tilde{w}},E)=H_{\rm min}(\vec{W}|h(\vec{W})=\vec{\tilde{w}},E).
\label{eq:H_min_fZ_equal_Hmin Z}
\end{equation}

The right hand side of (\ref{eq:H_min_fZ_equal_Hmin Z}) can be evaluated using the parity symmetry (\ref{eq:rho_z_E_parity_invariance}).
Let $s(\vec{\tilde{w}})$ be the number of `single' symbols included in $\vec{\tilde{w}}$ ({\it i.e.}, the number of single events), then there are $2^{s(\vec{\tilde{w}})}$ values of $\vec{w}$ satisfying $h(\vec{w})=\vec{\tilde{w}}$.
Because of (\ref{eq:rho_z_E_parity_invariance}), Eve's (sub-normalized) states $\rho^{\vec{\tilde{w}}}_E$ are equal for all these values of $\vec{\tilde{w}}$,
and thus the corresponding entropy takes the value
\begin{equation}
H_{\rm min}(\vec{W}|h(\vec{W})=\vec{\tilde{w}},E)= s(\vec{\tilde{w}}).
\label{eq:H_min_Wt}
\end{equation}
Finally, combining eqs. (\ref{eq:H_min_fZ_Hmin_fZT}), (\ref{eq:max_tH_min}), (\ref{eq:H_min_fZ_equal_Hmin Z}), and (\ref{eq:H_min_Wt}) together, we obtain
\begin{equation}
H_{\rm min}(g(\vec{W})|E)\ge \min_{\vec{\tilde{w}}} s(\vec{\tilde{w}}).
\label{eq:minst}
\end{equation}

\subsection{Equivalence of the ideal situation and the single photon RNG}
If we restrict ourselves with the ideal situation of Section \ref{sec:ideal_situation}, we can also show the security of our radioactive RNG by using the argument of complementary bases, which was mentioned in the eighth paragraph of Section \ref{sec:introduction} to show the security of single photon RNGs.
To see this, let $E_A:=E^\uparrow_A-E^\downarrow_A$.
Then because $E_A^2=P_A^2=\II_A$, Jordan lemma can be applied to $E_A$ and $P_A$.
Further, due to condition (\ref{eq:covariant_E}), we can decompose the Hilbert space ${\cal H}_{A}$ as ${\cal H}_{A}={\cal H}_{A_1}\otimes{\cal H}_{A_2}$ such that
\begin{equation}
E_A=\sigma^z_{A_1}\otimes \II_{A_2},\ P_A=\sigma^x_{A_1}\otimes \II_{A_2},
\end{equation}
where $\sigma^z,\sigma^x$ are the Pauli matrices.
Hence measurements of radiation directions $w_i$ and of parity becomes mathematically equivalent to those of $+$ and $\times$ bases used in the single photon RNG.
And one can prove the security of $w_i$ by using the same argument as in the seventh paragraph of Introduction.

\noindent{\bf Acknowledgments}
TS is supported in part by Cross-ministerial Strategic Innovation Promotion Program (SIP) (Council for Science, Technology and Innovation (CSTI)); CREST (Japan Science and Technology Agency) JPMJCR1671; JSPS KAKENHI Grant Number JP18K13469. TS thanks Quantaglion Co. Ltd. for useful information and discussion about the actual implementation of the radioactive RNG.

\bibliography{qrng_parity}

\end{document}